\documentclass[a4paper,fleqn,usenatbib]{mnras}

\usepackage[T1]{fontenc}
\usepackage{ae,aecompl}

\usepackage{graphicx}	% Including figure files
\usepackage{amsmath}	% Advanced maths commands
\usepackage{amssymb}	% Extra maths symbols

\title[Resolved pre-maximum halt in the nova V5589 Sgr]{Temporal resolution of a pre-maximum halt in a Classical Nova: V5589~Sgr observed with \textit{STEREO} HI-1B}

\author[S. P. S. Eyres et al.]{S. P. S. Eyres,$^{1}$\thanks{E-mail: spseyres@uclan.ac.uk (SPSE)}
D. Bewsher,$^{1}$
Y. Hillman,$^{2}$
D. L. Holdsworth,$^{1}$
M. T. Rushton,$^{3,1}$
\newauthor
D. Bresnahan,$^{1}$
A. Evans,$^{4}$
P. Mr{\' o}z$^{5}$
\\
$^{1}$Jeremiah Horrocks Institute, University of Central Lancashire,
Preston PR1 2HE, UK\\
$^{2}$Department of Geosciences, Raymond and Beverly Sackler Faculty
of Exact Sciences, Tel-Aviv University,
Tel-Aviv 69978, Israel\\
$^{3}$Astronomical Institute of the Romanian Academy, Str. Cutitul de
Argint 5, 040557, Bucuresti, Romania\\
$^{4}$Astrophysics Group, Keele University, Keele, Staffordshire ST5 5BG, UK\\
$^{5}$Warsaw University Observatory, Al. Ujazdowskie 4, 00-478 Warszawa, Poland
}

\date{Accepted 2017 January 31. Received 2017 January 26; in original
  form 2016 December 22}

\pubyear{2017}

\begin{document}
\label{firstpage}
\pagerange{\pageref{firstpage}--\pageref{lastpage}}
\maketitle

\begin{abstract}
Classical novae show a rapid rise in optical
  brightness over a few hours. Until recently the rise phase,
  particularly the phenomenon of a pre-maximum halt, was
  observed sporadically. Solar observation satellites
  observing Coronal Mass Ejections enable us to observe the
  pre-maximum phase in unprecedented temporal resolution. We present
  observations of V5589~Sgr with \textit{STEREO} HI-1B at a cadence of
   40~min, the highest to date. We temporally resolve a
    pre-maximum halt for the first time, with two examples each rising over
    40~min then declining within 80~min. Comparison with a grid of outburst models
  suggests this double peak, and the overall rise timescale,
  are consistent with a white dwarf mass, central temperature and
  accretion rate close to $1.0~\mathrm{M}_\odot$, $5 \times
  10^7~\mathrm{K}$ and $10^{-10}~\mathrm{M}_\odot ~\mathrm{yr}^{-1}$
  respectively. The modelling formally predicts mass loss onset at JD~2456038.2391$\pm$0.0139, 12~hrs before
  optical maximum. The model assumes a main--sequence donor.
  Observational evidence is for a subgiant companion; meaning the accretion rate is under--estimated. Post--maximum we see
  erratic variations commonly associated with much slower
  novae. Estimating the decline rate difficult, but we
  place the time to decline two magnitudes as $2.1 < \mathrm{t}_2
  (days) < 3.9$ making V5589~Sgr a ``very fast''
  nova. The brightest point defines ``day 0'' as
  JD~2456038.8224$\pm$0.0139, although at this high cadence the meaning
  of the observed maximum becomes difficult to define. We suggest that
  such erratic variability normally goes undetected in faster novae
  due to the low cadence of typical observations; implying erratic behaviour is not necessarily related to the rate of decline.
\end{abstract}


%%%%%%%%%%%%%%%%%%%%%%%%%%%%%%%%%%%%%%%%%%%%%%%%%%

%%%%%%%%%%%%%%%%% BODY OF PAPER %%%%%%%%%%%%%%%%%%

\section{Introduction}
\label{sec:intro}

Classical novae are interacting binary stars, in which a white dwarf
(WD) accretes matter from a (usually) main--sequence--like companion,
generally in a close orbit with a period of a few hours to days. This
results in a build up of material on the WD surface, leading to
thermonuclear runaways (TNRs) that cause an outburst, at ultraviolet
(UV), optical and infrared (IR) wavelengths in a few hours, and the
ejection of around $10^{-5}$ to $10^{-4} \mathrm{M}_\odot$ of matter
at velocities of a few $100$ to a few $1000~\mathrm{km}~\mathrm{s}^{-1}$
\citep{Warner2008}. Decline then follows over the course of weeks or
months, with the time to drop 2 or 3 magnitudes from maximum related
to the absolute magnitude via the maximum-magnitude-rate-of-decline
relationship \citep[MMRD;][]{Downes2000}, for both Galactic and
extra--galactic populations \citep{Darnley2006}. Slower novae often
show erratic variations and remain bright for longer times than is
consistent for the MMRD, when considered in the light of independent
distance measurements. The relationship is also not useful for novae
that decline most rapidly, as the absolute magnitude becomes
insensitive to decline time \citep[e.g.][]{Capaccioli1990}.

Recently observations and theoretical developments have established
that, prior to optical maximum X--ray and even gamma ray emission is
seen \citep{Ackerman2014}. Due to the rate of ascent, the {\em
  pre--maximum} optical light--curve has been only rarely observed, and
at cadences of $1-2$ data points per day. Some show a pre--maximum
halt (PMH) that had not been subjected to theoretical
modelling. 

\citet{Hounsell2010} presented four light--curves taken with
the Solar Mass Ejection Imager (\textit{SMEI}) down to time resolutions of
80~min, showing the PMH to be ubiquitous in their
sample. Further analysis of SMEI light-curves of novae was presented by \citet{Hounsell2016}, with a catalogue of 14 systems. They derived peak time, maximum magnitude and an estimate of the rate of decline for all systems, and tentatively detected PMHs in two of their sample.\citet{Hillman2014} {\bf have} demonstrated that the form of these
halts varies with the gross parameters of WD mass, WD central
temperature, and the accretion rate from the companion.

Here we present data from the \textit{STEREO} HI-1B camera for
V5589~Sgr, following the outburst from the initial rise, through the
optical peak to the point where it fades below detectability for this
instrument. The cadence is around 40~min, making these the highest
time-resolution observation of this phase to date. We discuss the rise
phase, including the PMHs, in the context of the models of
\citet{Hillman2014}. We also contrast the post--maximum behaviour with
much slower Classical Novae (CNe) and suggest that the empirical split
in this behaviour between slow and fast novae is in part a function of
the typical one day sampling of the light--curves.

\section{V5589 Sgr}
\label{sec:v5589sgr}

V5589~Sgr (Nova~Sgr~2012, PNV~J17452791--2305213, coordinates J2000
17~45~28.03 $-$23~05~22.7) was discovered on 2012~April~21.0112~UT
\citep{Sokolovsky2012} at $\mathrm{V} \sim 9.6$, rising to a reported
maximum of $\mathrm{V} = 8.8$ \citep{Korotkiy2012} by
2012~April~21.654~UT \citep{CBAT2015}. Pre--discovery images detected
the outburst on 2012~April~20.8403~UT down to $\mathrm{V} = 10.2$
\citep{CBAT2015}.

Early spectroscopy indicated high velocity ejecta in this nova, with
line widths (FWHM) of $5600~\mathrm{km}~\mathrm{s}^{-1}$
\citep{Korotkiy2012} to $6500~\mathrm{km}~\mathrm{s}^{-1}$
\citep{Esipov2012}. Observations in X-ray \citep{Sokolovsky2012,
  Nelson2012a} and radio \citep{Nelson2012a} wavelengths shortly after
optical peak showed non--detections, and suggested no shocked
component in the expanding ejecta at this early stage of the
decline. 

Swift UVOT detected the object on 2012~April~25.7 at a UVM2
brightness of $13.71 \pm 0.09$~mag, fading rapidly from $13.8$ to
$13.5$ over a 5975\,s observation \citep{Nelson2012b}. By 2012~May~10
hard X--ray emission was present, consistent with a column of $N(H) =
(3 \pm 1) \times 10^{21}$~cm$^{-2}$ and a plasma temperature of at
least $2.7 \times 10^8$~K; the UVM2 brightness was $16.0 \pm 0.2$~mag
by this date \citep{Nelson2012b}. \citet{Weston2016} argue that the
radio and X-ray characteristics are consistent with high--speed,
low--mass ejecta that accelerates over the first few weeks after
outburst. 

The nova has been observed by the OGLE survey since 2010
\citep{Mroz2015}. In the quiescent light--curve they found eclipses
with a period of 1.59230(5)~days, which they interpret as the orbital
period. Such a relatively long orbital period suggests that the
secondary is a subgiant. If the secondary fills its Roche lobe the
orbital period can be used to estimate the radius of the secondary
\citep[see e.g.][]{King1989}. Using Eq (2.3.10) of King we find that
the secondary radius $R_2/R_\odot \sim 0.11 \mbox{P}_{\rm hr}  \sim
4.2R_\odot$, confirming that the secondary is evolved.  
The last pre-eruption image was taken on 2012 April 20.3728~UT at $I =
16.90$. These authors also note that with this longer period, the
companion is likely to be a subgiant. \citet{Weston2016} come to the
same conclusion.

\section{Observations}
\label{sec:observations}

The Solar TErrestrial RElations Observatory (\textit{STEREO}) consists
of two nearly identical spacecraft, in heliocentric orbits, one ahead
of the Earth (STEREO-A) and the other behind the Earth (STEREO-B)
\citep{Kaiser2008}. The twin spacecraft were launched on 2006 October
26, and the separation between each spacecraft and the Earth has
increased by approximately 22$^{\circ}$ per year since achieving
helocentric orbits.

The Sun Earth Connection Coronal and Heliospheric Investigation
({\sc secchi}) instrument suite \citep{Howard2008} on \textit{STEREO}
includes the heliospheric imagers (HI-1 and HI-2). The HI instruments
have been used to observe a variety of  variable stars including
classical nova V5583~Sgr \citep{Holdsworth2014}, and long period variables
\citep{Wraight2012}.

Data from the HI-1 instument on \textit{STEREO-B} (HI-1B) are used in
this paper. HI-1B has a field-of-view (FOV) of $20^\circ$, which
extends from $\sim 4 - 24^\circ$ elongation, with a pixel size of
$\sim 70$~arcsec, \citep{Eyles2009}. The \textit{STEREO} HI-1B
instrument's main passband covers the range from $600 - 750$nm, but
also has contributions at $300 - 450$nm and from $900 - 1100$nm
\citep[see Fig.~6 of][]{Bewsher2010}

The HI-1B images consist of 30 exposures of 40\,s each, with a
summed image cadence of 40~min \citep{Eyles2009}. Each exposure is
scrubbed of cosmic ray hits prior to summing \citep{Howard2008}. The
summed images constitute the L0.5 {\sc fits} files which are available to
download from several websites including the UK Solar System Data
Centre (UKSSDC\footnote{http://www.ukssdc.ac.uk/}).

The L0.5 data have been prepared with the {\sc sswidl} procedure,
{\sc secchi\_prep}. This routine applies the shutterless correction to the
data, the large scale flatfield determined by \citet{Bewsher2010}, and
the pointing and optical parameter calibration of
\citet{Brown2009}. It also identifies missing blocks and saturated
pixels, filling these pixels with NaN values.

As discussed in \citet{Bewsher2010}, the point spread function (PSF)
of the HI-1B instrument is well approximated with a Gaussian profile,
and the aperture size of 3.1 pixels suggested by \citet{Bewsher2010}
is used to extract the nova light--curve from the data using aperture
photometry. The solar F-corona is the dominant source in the HI-1B
data, and is removed using a daily running background.

To extract the light--curve presented in this paper, each HI image is
processed with {\sc secchi\_prep}, and the position of the nova determined
using the pointing information given in the header of the {\sc fits}
file. Aperture photometry is performed on this position, and the
number of counts (DN/s) recorded. This process is repeated for every
HI image from 2012~April~18 to 2012~May~7, where the nova crosses the
CCD of the instrument. The counts recorded are then converted into a
magnitude using the following equation
\begin{equation}
M_{HI} = -2.5 \log \left( I_{tot} \over \mu F \right)
\end{equation}
where $M_{HI}$ is the HI magnitude, $I_{tot}$ is the recorded counts,
$\mu$ is a calibration conversion factor and $F$ is the flux of Vega
\citep{Bewsher2010}, the zero--point of the magnitude scale. The error
on $\mu F$, $\sigma_F$, is estimated to be of the order of 5\%,
therefore,
\begin{equation}
{\sigma_F \over \mu F} = 0.05
\end{equation}

It is assumed that the uncertainties on $I_{tot}$, $\sigma_I$ are
Poisson counting statistics, and the conversion from DN (units of
$I_{tot}$) to photoelectrons is $1~\mathrm{DN} =
15~\mbox{photoelectrons}$ \citep{Eyles2009}, therefore,
\begin{equation}
\sigma_I = \sqrt{I_{tot} \over 15}.             
\end{equation} 

The uncertainty on the magnitude measurement is then given by
\begin{eqnarray}
\sigma_{M_{HI}} & = & \left| {-2.5 \over \ln(10)} \sqrt{ \left({\sigma_I \over I_{tot}}\right)^2 + \left({\sigma_F \over \mu F}\right)^2} \right|\\
& = & \left| {-2.5 \over \ln(10)} \sqrt{ \left({\sqrt{I_{tot} \over 15} \over I_{tot}}\right)^2 + \left(0.05\right)^2} \right|
\end{eqnarray}

This value was calculated on a point--by--point basis and is included in
Fig.~\ref{fig:lc} as error bars. The error due to the background
subtraction has not been included in the 
errors on the light--curve. However, it is estimated that the error due to
the background subtraction is of the order of 10\% of the calculated error
for a handful of points, and less than 5\% for the majority of points. This
is due to the background subtraction being calculated using the average
over a days worth of observations leading to comparatively small
uncertainties.

In the following we use the 40~min cadence of the data as the basis
for uncertainties in any dates or intervals in time. This translates
to $\pm$0.0139~d, with the third decimal place being conservative given that
observations are made at this cadence to within a few seconds. 

\section{Results}
\label{sec:results}

\begin{figure*}

	\includegraphics[width=\textwidth]{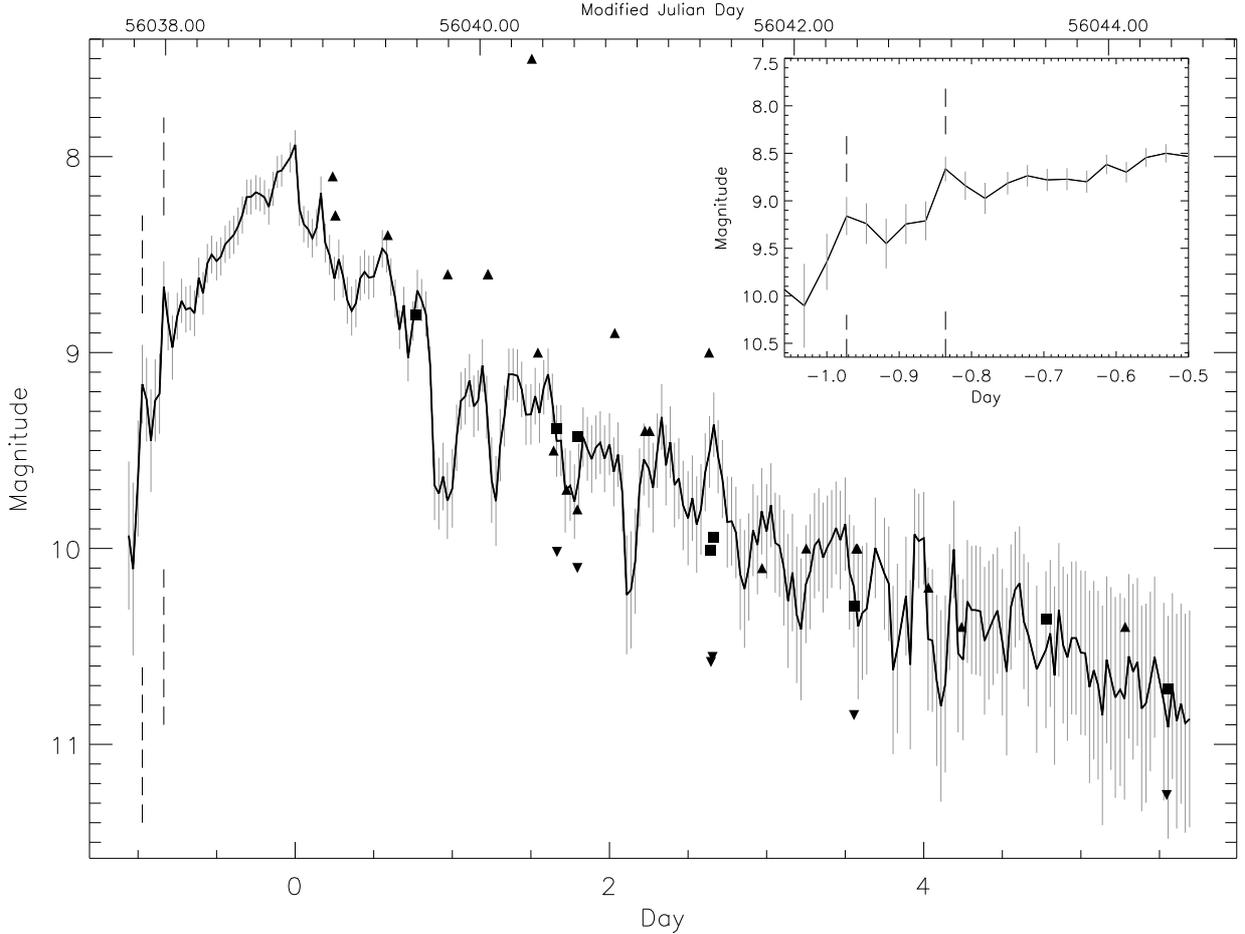}
        \caption{\textit{STEREO} HI-1B data with error bars, plus AAVSO
          visual estimates (triangles), V (squares) and B
          (inverted triangles) included for comparison; offset by 1~mag upwards
          for clarity. \textit{STEREO} is approximately R-band, for which data
          are available in AAVSO only after JD~2456051 (MJD 56051.5),
          about 6~d after the end of this data run. Vertical dashed
          lines mark the PMHs described in the text. The inset zooms
          in on the $\sim$14~hour period around the PMHs.}
    \label{fig:lc}
\end{figure*}

The light--curve of V5589 Sgr is presented in Fig.~\ref{fig:lc}. We
see the rise to maximum, and then a rather erratic decline. We also
overlay American Association of Variable Star Observers (AAVSO) B, V and visual estimate data for comparison. Uncertainties
in AAVSO visual estimates are estimated as 0.1~mag although this varies with
location and experience of the contributing observers. We can see that
while the \textit{STEREO} and AAVSO data are consistent, the latter has a much
lower cadence and relies primarily on visual estimates.

The time of maximum is a key characteristic determined for previous
novae, and often used to refer to subsequent development, with the
brightest point labeled as ``day 0''. As the decline time from peak is
used to determine the speed class of CNe \citep{Payne-Gaposhkin1957,
  Duerbeck2008}, which in turn correlates with the energetics of the
event, it is useful to identify this point. There is one brightest
point in the \textit{STEREO} data, albeit less than 1~$\sigma$ above adjacent
points. This point is at JD~2456038.8224$\pm$0.0139, and can be taken
as the peak magnitude. 

Determining the time to decline by two magnitudes, $\mathrm{t}_2$, and
hence the speed class, is somewhat more uncertain. The brightness at
peak is $7.94 \pm 0.09$, and the brightness drops below 9.94 for the
first time after JD~2456040.9057$\pm$0.0139 but recovers 2~h
later. The last time it is brighter than this is the point at
JD~2456042.3224. The average magnitude binned over 18 points (12~h interval)
reaches this level in the bin centred on JD~2456041.7668. These give a
$\mathrm{t}_2$ time of between 2.1 and 3.9~days, placing the nova
firmly in the ``very fast'' class. The \textit{STEREO} passband incorporated
more red light than the B or V bands where $\mathrm{t}_2$ is usually
determined. Nonetheless the values are well within the ``very fast''
range, and there is an expectation that this object will be in the
upper range for intrinsic luminosity and velocity of the ejecta of the
nova events. This is consistent with the identification of a
high--speed shock by \citet{Weston2016}.

\subsection{Pre--maximum halts}
\label{sec:premax}

A key finding of our work is the presence in the rise--phase of
deviations from a monotonic increase that are consistent with PMHs
seen in other classical novae. For V5589~Sgr we have identified two
short deviations -- peaking at JD~2456037.8502$\pm$0.0139 and
JD~2456037.9891$\pm$0.0139 -- with 
this phenomenon. From Fig.~\ref{fig:lc} we can see that the rise to
peak occurs between one point and the next (i.e. within 40~min), with
the decline over three points (i.e. over 80~min), both with the same
shape within the uncertainties. As the PMHs appear V5589~Sgr moves
between pixels 80 and 110 along the 540th row of the CCD, so the nova
is well away from the edge of the 1024$\times$1024 pixel detector. 

The initial rise reflects a change of more than $3~\sigma$
point-to-point as a PMH starts (uncertainties decline with increasing
brightness).  The PMH decline is to a level within $2~\sigma$ of the
pre--rise magnitude. Modelling by \citet{Hillman2014} suggests that
PMHs are often short--lived excursions above the general upwards trend
of the rise. We explore this by taking linear interpolations between
the ``before'' and ``after'' points for both features. These are found
to follow a line consistent with the subsequent rise. Thus the two
peaks are consistent with being such excursions from the general
upwards trend, rather than being defined by dips that interrupt the
rise. The combination of the change relative to the uncertainties and
the deviation from what one would expect from backwards extrapolation
of the rise, allows us to be confident that they can be reasonably
described as PMHs as modelled by \citet{Hillman2014}.

\section{Discussion}
\label{sec:discussion}

We discuss the pre-maximum halts during the rise phase, and
then the behaviour following the optical peak.

\subsection{Double pre--maximum halt}
\label{sec:doublehalts}

During the rise to maximum (Fig.~\ref{fig:lc}), we see two very
similar features that represent departures from the smooth, monotonic
rise normally considered to be the usual behaviour for CNe before the
optical peak. We have interpreted these as the sometimes--observed
PMHs. When taken in combination with those seen by
\citet{Hounsell2010} we can conclude that these events are common, and
have been generally missed in historical observations.

\begin{figure}
	\includegraphics[width=\columnwidth]{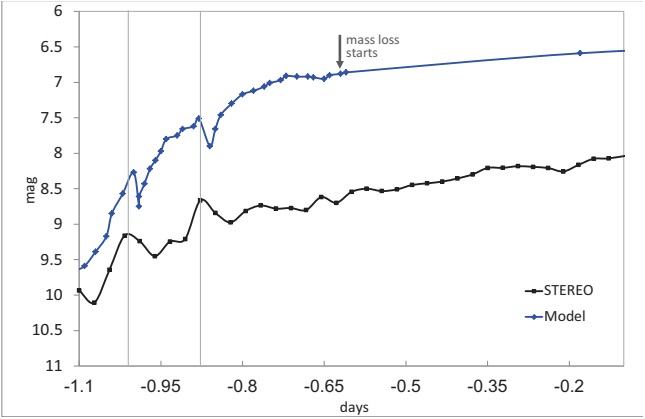}
        \caption{Comparison of data (lower line) and model (upper
          line) for the parameters giving the closest match to the
          pre--maximum behaviour (WD mass $1.0~\mathrm{M}_\odot$,
          central temperature $5 \times 10^7~\mathrm{K}$ and mass accretion rate
          $10^{-10}~\mathrm{M}_\odot~\mathrm{yr}^{-1}$). The time axis is relative
          to day~0 at JD~2456038.8224.}
    \label{fig:model}
\end{figure}

In Fig.~\ref{fig:model} we compare a model from the grid of
\citet{Hillman2014} with the \textit{STEREO} data. We have chosen
parameters which give the best qualitative agreement with the data. We
have not attempted to fit the parameters to the data, but instead we
offer this as support that the general development of the PMHs during
the rise is understood. The model plot shows the relevant time period
from Fig.~4, Panel~4, plot 100.50.10 of \citet{Hillman2014}. The
model starts from well before the span of the data, and to assess the
similarity, we have simply aligned the peak of the second halt. By
inspection we can then see that the duration, magnitude and interval
between the halts is consistent between the model and the data. The
PMHs occur in the model at the point when convection ceases to be
efficient near the surface causing a reduction in the energy flux. The
opacity then decreases allowing radiation to become the dominant
energy transfer mechanism and reverse the dip. This starts to drive
mass loss.

The model agrees less well with the behaviour of V5589~Sgr once
convection ceases to dominate evolution around the PMHs. Nonetheless we
have constrained the onset of mass loss to occur between the peak of
the first PMH and the maximum of the optical light (i.e. between
JD~2456037.8502 and JD~2456038.8224), with the model formally giving
this as JD~~2456038.2391. The modelling assumes a main--sequence donor
star. While the subgiant donor would have a lower H abundance, this
simply means the accretion rate derived from the model is an
underestimate of the true rate. An optical spectrum in quiescence
would be valuable to confirm the nature of the secondary.

\subsection{Post--maximum erratic decline}
\label{sec:postmax}

There are a number of excursions visible in Fig.~\ref{fig:lc}
signifcantly above or below the mean magnitude during the decline
after day~0. We consider if any of this structure is consistent with
periodicity. We also examine how significant the excursions are given
the overall scatter in the data.

An orbital period of 1.59230~d determined by \citet{Mroz2015} is
based on the ephemeris of $\mathrm{HJD} = 2455000.724 \pm 0.015 +
(1.59230 \pm 0.00005) \times \mathrm{E}$ for the eclipses.

To test for periodicities in the decline phase, we excluded data
obtained before day 0 and perform a discrete Fourier transform on the remaining
data using the {\sc{Period04}} software of
\citet{2005CoAst.146...53L}. After fitting and removing the slope of
the decline, we search for significant periods in the de-trended
data. Given the quality of the observations, we relax the `standard'
S/N limit of 4 \citep[e.g.][]{Koen2010} to 3 to determine
if a peak in the FT is significant, however we find no peaks in the FT
above this limit. Furthermore, there is no evidence in our data of the
orbital period determined by \citet{Mroz2015}.

This is consistent with post--outburst Optical Gravitational Lensing
Experiment (OGLE) observations showing that eclipses do not start
occuring again until HJD = 2456062.82544 or possibly as late as HJD =
2456076.92252. Prior to this the entire binary must be within the
photosphere of the nova ejecta, and so must be larger than the orbital
radius of $\sim 7.2R_\odot\times(M/2M_\odot)^{1/3}$ for a system mass
of $M$.

While there are a number of variations in the decline phase, none are
consistently separated by this interval, or a multiple thereof. In
addition a simple polynomial fit to the decline suggests only the
minima centred around MJD 56039.7668, 56040.1002 and 56040.9335 are
consistent with anything other than the overall level of fluctuations
seen. None of these are separated by the period in
\citet{Mroz2015}. Thus whatever the origin of these dips, neither
eclipses nor orbital modulation appear to explain them.

The speed class appears to be related both to the luminosity of the
nova at the peak, and the kinetic energy of the ejecta, with both
being higher for faster novae. Much less consistently, more slowly
declining examples often show erratic brightening and dimming after
the initial rise, presenting multiple false maxima and making the
determination of the peak, and hence the decline rate,
difficult. \citet{Evans2003} discuss V723~Cas as an example,
and suggest these ``flare-like'' features might be due to interactions
within the ejecta or mass transfer bursts from the companion, as
proposed by \citet{Chochol2000}. By contrast erratic behaviour is
not generally seen in fast examples of CNe, where the rise and fall is
seen to be smooth. However the vast majority of ground--based
light--curves are made up of a number of observations every night,
with considerable variation in the consistency of observations and are
mainly visual estimates. The data presented here provide unprecedented
stability and point-to-point reliability at a regular 40~min
cadence.

\citet{Hounsell2010} found some erratic behaviour early on in
V1280~Sco and to a lesser extent in KT~Eri, the second also being very
fast. As noted in their Section~4 the decline time and the form of the
PMHs do not seem to have any correlation; V5589~Sgr shows a different
form again even though the initial decline rate was similar to that of
KT~Eri. We suggest that the association of erratic brightness
variations near peak with slower novae is a feature of the poor
cadence of most light--curve data to date. It is clearly a feature of
some examples of fast novae, but at this point we do not have enough
examples to judge if it is less common than in slower
novae. \citet{Hounsell2016} showed further examples of non-smooth
declines from maximum in many of their novae light curves, however due
to the limiting magnitude and contamination the presence of erratic
variations may not be astrophysical.

In fact, once we start to record the early light--curve of novae at the
cadence achieved here, the definition of ``day 0'' needs to be
revisited. An alternative approach to determining the peak in
Fig.~\ref{fig:lc} is to treat all fluctations from a smooth rise and
decline as anomalous. In this case some form of fit to the overall rise
and decline might be used to give an estimate of the peak that is
derived in a manner consistent with those for novae with only low
cadence data. A second--order polynomial fit through the peak has a
maximum of $8.16 \pm 0.10~\mathrm{mag}$, at 56038.68, or 160~min
earlier than the absolute brightest measurement used to define ``day
0'' in Section~\ref{sec:results}. Thus at this high cadence, where we
see fluctations on timescales of a few hours, the relevance of optical
maximum and hence ``day 0'' starts to become problematic. However as
\textit{STEREO} and other facilities allow us to examine an increasing number
of fast novae in this detail, we may be able to reassess the speed
class concept.

\section{Conclusions}

We have collected an optical light--curve of the fast Classical Nova
V5589~Sgr at a time--resolution of 40~min, the best for any nova to
date. This data show the rise to maximum with pre--maximum halts, and
the subsequent somewhat erratic decline, in greater detail than seen
before. We find that the halts are consistent with a binary model
having a WD of mass $1.0~\mathrm{M}_\odot$, central temperature $5
\times 10 ^7~\mathrm{K}$ and mass accretion rate
$10^{-10}~\mathrm{M}_\odot~\mathrm{yr}^{-1}$. As the donor star is a
subgiant, the last of these is likely an underestimate. The model
shows the PMHs are pre--cursors to the onset of mass loss and formally
predicts this as starting at JD~2456038.2391$\pm$0.0139 for
V5589~Sgr.

We suggest that the erratic decline behaviour, a common characteristic
of slow novae, is missed in previous observations of some faster
examples by the typical daily observations. Our work, and that of
\citet{Hounsell2010}, demonstrate that some fast novae also demonstrate
erratic behaviour close to optical peak. This is not understood or
well modelled at this time. We also find that at such high cadence,
the common identification of the peak as ``day 0'' becomes
increasingly difficult and possibly misleading.

\textit{STEREO--A} continues to operate and we will retrieve data of
further CNe that are serendipitously bright as they cross one or more
of the four camera apertures. This will allow us to investigate the
prevalence and nature of PMHs, and any erratic behaviour close to
optical maximum.

\section*{Acknowledgements}

Data created during this research is openly available from the
University of Central Lancashire data repository, UCLanData, at
http://doi.org/10.17030/uclan.data.00000049.

The Heliospheric Imager (HI) instrument was developed by a
collaboration that included the Rutherford Appleton Laboratory and the
University of Birmingham, both in the United Kingdom, and the Centre
Spatial de Li\'{e}ge (CSL), Belgium, and the US Naval Research
Laboratory (NRL),Washington DC, USA. The \textit{STEREO}/{\sc secchi}
project is an international consortium of the Naval Research
Laboratory (USA), Lockheed Martin Solar and Astrophysics Lab (USA),
NASA Goddard Space Flight Center (USA), Rutherford Appleton Laboratory
(UK), University of Birmingham (UK), Max-Planck-Institut f\"{u}r
Sonnensystemforschung (Germany), Centre Spatial de Li\`{e}ge
(Belgium), Institut d'Optique Th\'{e}orique et Appliqu\'{e}e (France),
and Institut d'Astrophysique Spatiale (France).

DLH acknowledges financial support from the STFC via grant ST/M000877/1.

%%%%%%%%%%%%%%%%%%%%%%%%%%%%%%%%%%%%%%%%%%%%%%%%%%



%\bibliographystyle{mnras}


%%%%%%%%%%%%%%%%%%%%%%%%%%%%%%%%%%%%%%%%%%%%%%%%%%

%%%%%%%%%%%%%%%%% APPENDICES %%%%%%%%%%%%%%%%%%%%%

%%%%%%%%%%%%%%%%%%%%%%%%%%%%%%%%%%%%%%%%%%%%%%%%%%

% Don't change these lines
\bsp	% typesetting comment
\label{lastpage}
\end{document}